\documentclass{ws-ijmpc}

\usepackage{graphicx}

\begin{document}

\markboth{Veit Schw\"ammle}
{Simulation for competition of languages with an ageing sexual population}

\catchline{}{}{}{}{}

\title{Simulation for competition of languages with an ageing sexual population}

\author{V.~Schw\"ammle}
\address{Instituto de F\'isica, Universidade Federal Fluminense, Av. Litor\^anea, s/n - Boa Viagem, 24210-340, Niter\'oi, RJ, Brasil.\\
Institute for Computer Physics, University of Stuttgart, Pfaffenwaldring 27, D-70569 Stuttgart, Germany.\\veit\@if.uff.br}

\maketitle

\begin{history}
\received{}
\revised{}
\end{history}



%

\begin{abstract}
Recently, individual-based models originally used for biological purposes revealed 
interesting insights into processes of the competition of languages. 
Within this new field of population dynamics a model considering sexual populations
with ageing is presented. 
The agents are situated on a lattice and each one speaks one of two languages or both.
The stability and quantitative structure of an interface between two regions, initially speaking 
different languages, is studied. We find that individuals speaking both languages do not 
prefer any of these regions and have a different age structure than individuals speaking 
only one language.
\end{abstract}

\ccode{PACS Nos.:89.75.-k,89.65.-s}
  
\section{Introduction}
\label{sec:intro}

Physicists investigate biological and sociological systems, applying their tools, 
borrowed mainly from statistical physics. In the last years, the evolution of languages 
and competition among them gained much interest of non-linguistics, especially researchers 
working
in evolutionary systems similar to biological ones. There have been several approaches to reveal the 
similarities between the evolution of biological systems and languages at the end of the 
nineteenth century~\cite{Science2004}.
Numerical models simulating the competition of many languages have given new insights into the 
behavior of agents, for instance on a 
lattice~\cite{Abrams2003,Patriarca2004,Schulze2005,Kosmidis2005}, as well as into the
size distribution of languages~\cite{Schulze2005b}. 
A review of them can be found in ref.~\cite{Schulze2005}. 

Our model is based on the well understood Penna ageing model~\cite{Penna95,Moss99} on a 
lattice~\cite{Makowiec2001,Sousa99,Schwaemmle2005} 
which provides us a possibility to model a sexually reproducing stable population. 
We simplify the model by defining languages as an integer number, that is they are
not composed of different words. Thus we are
able to avoid changes in the same language. The parents pass their language entirely 
to their offspring. In order to 
stabilize their distribution, languages can be forgotten during lifetime.

The study of an interface between originally different regions under different parameter sets
reveals certain characteristics of the geographical distribution of the languages
on the lattice as well as of the age structure of the agents for different languages.

This articles is organized as follows: the next section explains the main 
features of the model and tries to justify the parameters we use. 
We present the results and the conclusions in the two following sections. 

\section{The model}
\label{sec:model}

This section is separated into two subsections in order to provide the reader a small review 
of the Penna model on a lattice as well as to present our modifications adapting 
the model to a system where languages compete with each other. 

\subsection{The sexual Penna model on a square lattice}

Each individual or agent comprises two bit-strings (diploid) of 32 bits
that are read in parallel. 
After birth, at every time step a new position of both bit-strings is read. 
A bit equal to 1 corresponds to a harmful allele.
All individuals have five predefined dominant positions where one harmful allele
suffices
to represent an inherited disease starting to diminish health from that age on
which corresponds to the bit position. 
At the other positions two set bits are needed to switch on the effect of a disease.
As soon as the agent reaches an age at which the current 
number of deleterious mutations exceeds the threshold value $T=3$, the agent
dies. In order to have a stable population, every time step 
an individual dies with the additional probability: $V=N(t)/N_{max}$
where $N(t)$ is the actual population size and $N_{max}$ the so called 
carrying capacity representing a maximum population size. After reaching 
the minimum reproduction age $R=10$, a female agent searches every time step 
for a male, of age equal or greater than the minimum reproduction age, among the 
central site and its four nearest neighbors to generate two offspring. 
It selects a male on the central site with a probability of 25\%, and if it fails, 
it searches among the nearest neighbor sites, at each one with a probability of 
25\%. The two bit-strings 
of the offspring are built by a random crossing and recombination of the parents
bit-strings (see ref.~\cite{Moss99}). Each new bit-string suffers a deleterious 
mutation (from zero to one) at a random 
position. If the selected bit is already 1, it remains 1 in the offspring
bit-string. The offspring is placed on a nearest neighbor site of the mother 
even if the site is already occupied (which is different from the 
usual versions of the Penna model on a lattice). Every time step an agent 
moves to a randomly selected nearest neighbor site with probability
$p_m$, if this site is less or equally populated.
The bit-strings are initialized randomly with zeros and ones at the first time step.

For a more detailed description of the Penna model and its implementation on a 
square lattice we refer to refs.~\cite{Moss99,Makowiec2001}.

\subsection{Competition of language}

For simplicity we define a language by an integer number and not by a bit-string 
which would describe, for instance, different words or an alphabet
as in ref.~\cite{Schulze2005,Kosmidis2005}.

Every agent speaks a language $l$, an individual variable which can have 
three values: $l=1$ and $l=2$ mean that it speaks language 1 or 2, respectively. 
The third possibility, $l=3$, describes the case where an agent speaks both 
languages. Our model contains two parameters dealing directly with these language 
values. At birth an offspring learns the language $l$ of its parents if 
they speak the same one(s). In the case of parents with different values of
$l$ the offspring speaks both languages ($l=3$) with probability $p_b$, 
otherwise it speaks only one language, each with the same probability.
The other parameter is the probability $p_f$, for which an agent may forget an 
already learned language. Every time step an agent, 
which speaks {\em both} languages, counts 
the number of surrounding agents speaking language $l$ in its neighborhood. 
This neighborhood is defined by a square of a distance of $d$ sites from the central
 site, for instance the 8 nearest neighbors with $d=1$ or 24 with $d=2$. 
The central site is not counted. If and only if there is a majority of 
people in the neighborhood 
speaking language 1 or 2 the agent forgets language 2 or 1 with probability $p_f$, 
respectively. Thus it speaks only the language which dominates in its surrounding.
 The lattice has free boundary conditions.


\section{Results}
\label{sec:res}

In our simulations we restrict ourselves to the following initial conditions:
half of the lattice of size $L \times L$ is filled up
with agents speaking language $l=1$ and the other half with language $l=2$.
We study stability and shape of the interface between these two regions.
Simulations with randomly distributed languages do not present a stable interface:
The whole population speaks only one and the same language after a few time steps. 

The initial population consists of $10,000$ males and $10,000$ females 
randomly distributed over the lattice with the values of $l$ as described 
above. The carrying capacity is $N_{max}=1,000,000$ on a $20 \times 20$ square 
lattice.

\begin{figure}[htb]
  \begin{center}
    \includegraphics[width=0.8\textwidth]{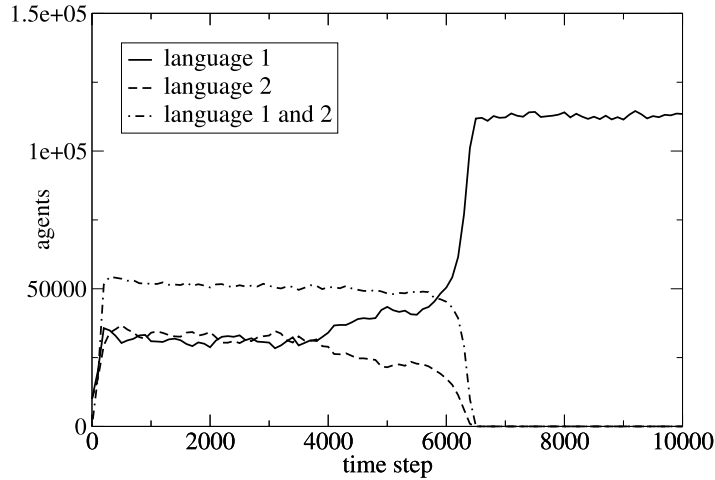}
\caption{Unstable interface: Only agents with $l=1$ remain.},
    \label{fig:inst}
\end{center}
\end{figure}

The simulations show that the interface is neither stable for $p_b$ values 
smaller than one nor for low 
occupation (less than 100 agents per site), which is controlled by the carrying
capacity $P_{max}$ and the lattice size. After a short time
the number of agents speaking $l$ begins to fluctuate strongly and finally
converges into the stable state where only one language is spoken.
In general the stability of the interface depends crucially on the initial 
state as well as on the random seed. For instance, Figure~\ref{fig:inst} 
shows a simulation with $p_f=0.1$, $p_b=1$, $p_m=0$ and $d=1$ presenting such an 
instability.

\begin{figure}[htb]
  \begin{center}
    \includegraphics[width=0.8\textwidth]{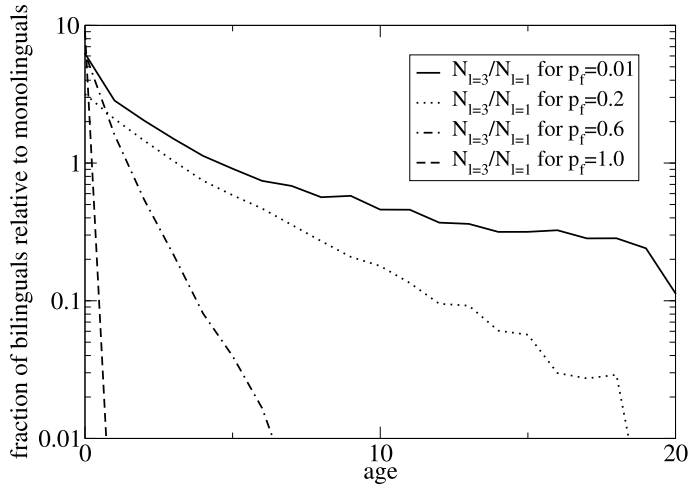}
\caption{The number of agents speaking language $l=3$ divided by the number
	 of agents with $l=1$, as a function of ages.
	 The number of agents speaking both languages decreases 
	 drastically with age for large values of $p_f$.}
    \label{fig:a_struct}
\end{center}
\end{figure}

\begin{figure}[htb]
  \begin{center}
    \includegraphics[width=0.8\textwidth]{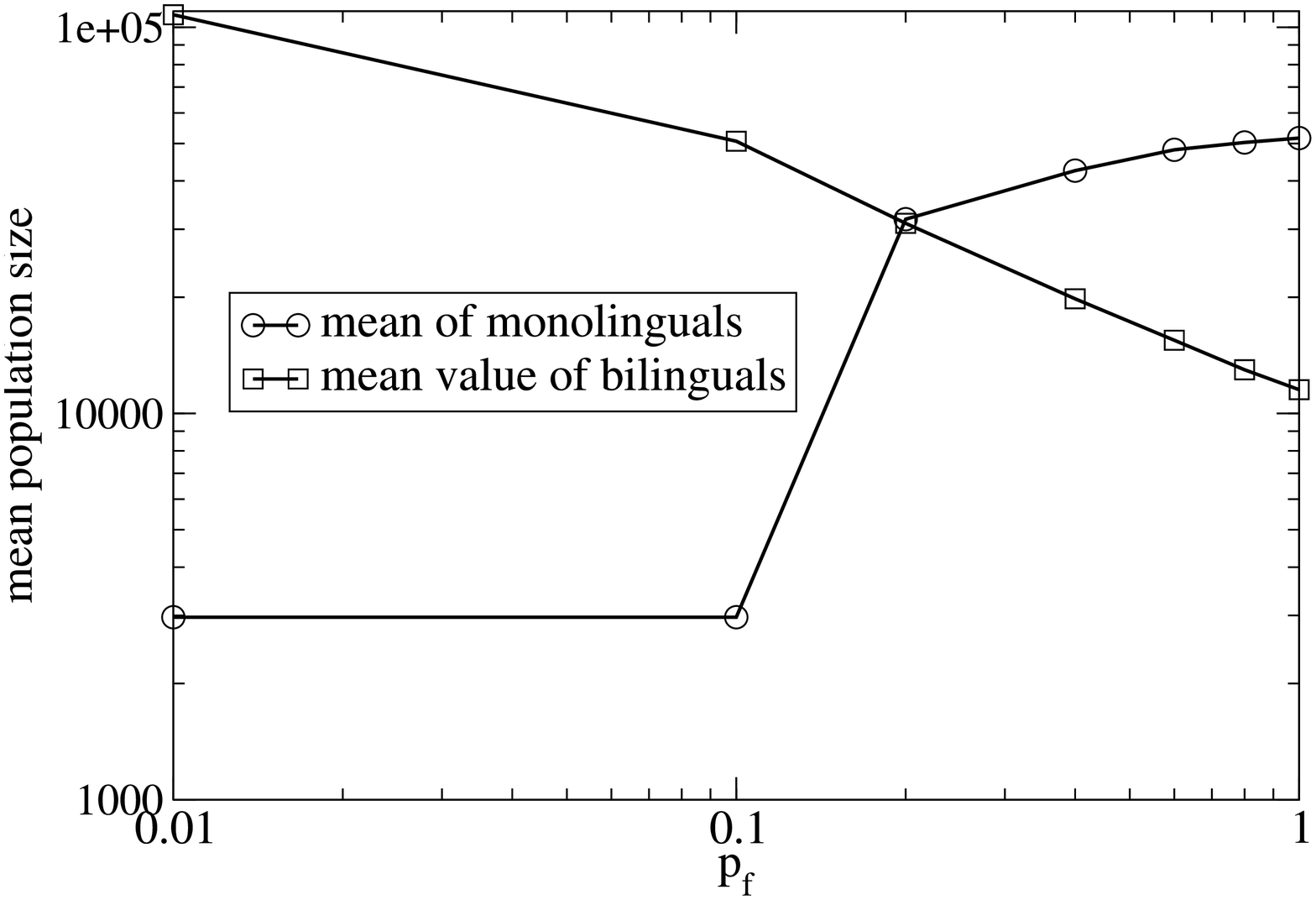}
\caption{The mean value of the number of bilinguals seems to decay as a power law.}
    \label{fig:freqPf}
\end{center}
\end{figure}

We concentrate now on the results of the simulations for different values of the 
parameter
$p_f$, leading an agent to forget one of its languages. We fix the other parameters
to $p_b=1$, $d=1$ and $p_m=0$. 
As a function of age, Figure~\ref{fig:a_struct} shows the population of bilinguals 
(agents speaking both languages) divided by the monolinguals ($l=1$ or $2$) for 
different values of $p_f$. The function decreases exponentially due to the 
effect that every time step a fraction of the bilinguals becomes monolingual. 
With increasing $p_f$ a larger fraction of older bilingual agents forgets 
one of their two languages since in their environment they do not need both.  
The higher is the probability to forget a language, the smaller is the number 
of older agents speaking both languages and thus less offspring with $l=3$ are 
created. Figure~\ref{fig:freqPf} depicts the mean value of monolinguals and 
bilinguals during one simulation for different values of $p_f$. The number of 
bilinguals decreases with increasing $p_f$ as expected. It seems that the 
fraction of bilinguals decays roughly as power law with exponent $-1$ for 
higher values of $p_f$.

\begin{figure}[htb]
  \begin{center}
    \includegraphics[width=0.8\textwidth]{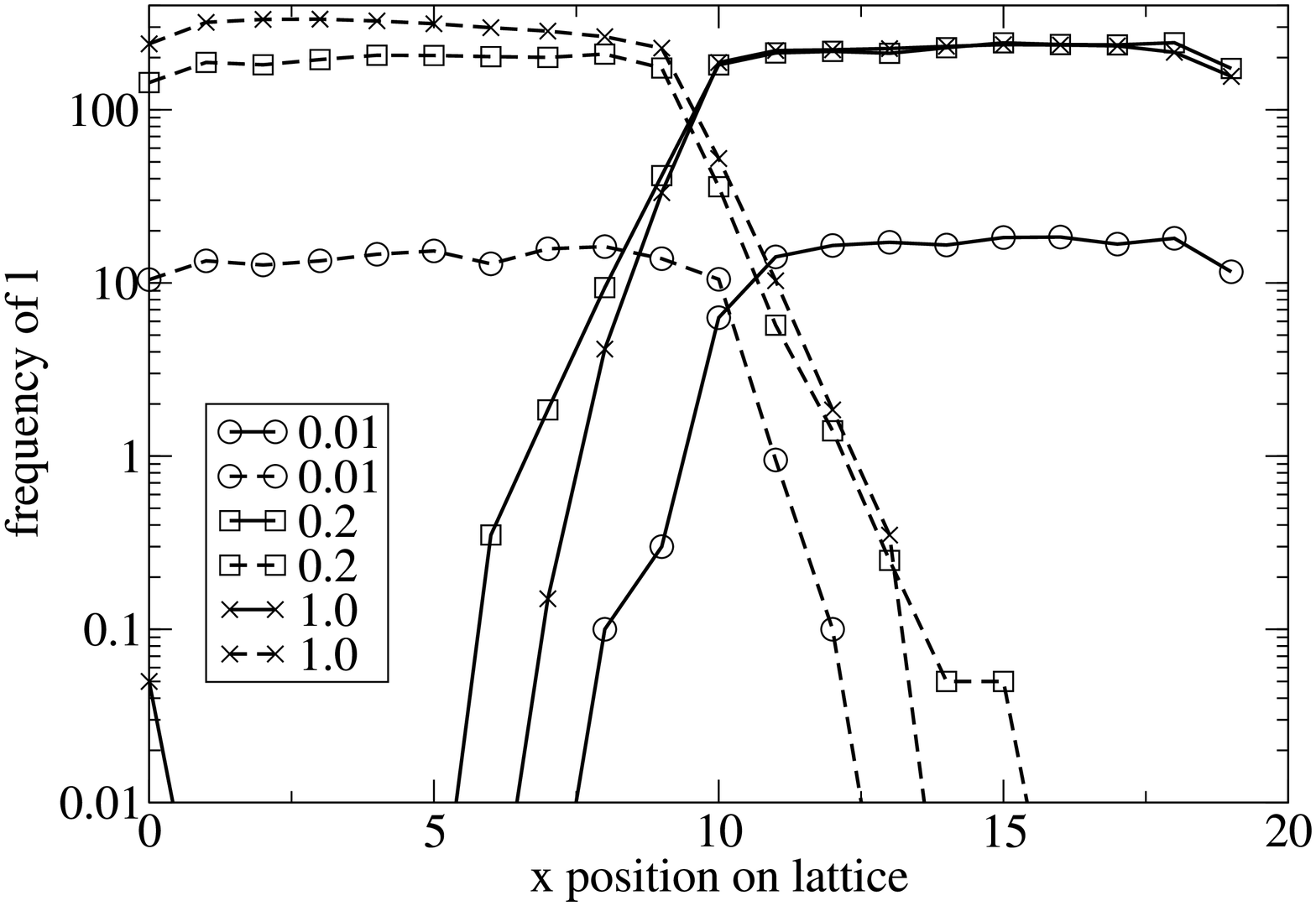}
\caption{The frequency of monolinguals saturates for high $p_f$ values. 
         The dashed line corresponds to $l=1$ and the solid line to $l=2$.
	 For all values of 
	 $p_f$ we find an exponential behavior at the interface.}
    \label{fig:distr_pf}
\end{center}
\end{figure}

Figure~\ref{fig:distr_pf} shows the number of agents with certain value of $l$ 
versus their position in direction $x$ perpendicular to the
interface. The number is averaged over the direction parallel to the interface 
and is measured after $10,000$ time steps. We observe a quite stable interface 
between the two regions, each one with one of the two languages in majority. 
The number of agents speaking $l=1$ and $l=2$
decays exponentially at the interface as also reported in ref.~\cite{Schulze2005}. 
Interestingly, 
the number of bilinguals is constant
over the whole lattice. The shape is not altered by changing $p_f$.

\begin{figure}[htb]
  \begin{center}
    \includegraphics[width=0.8\textwidth]{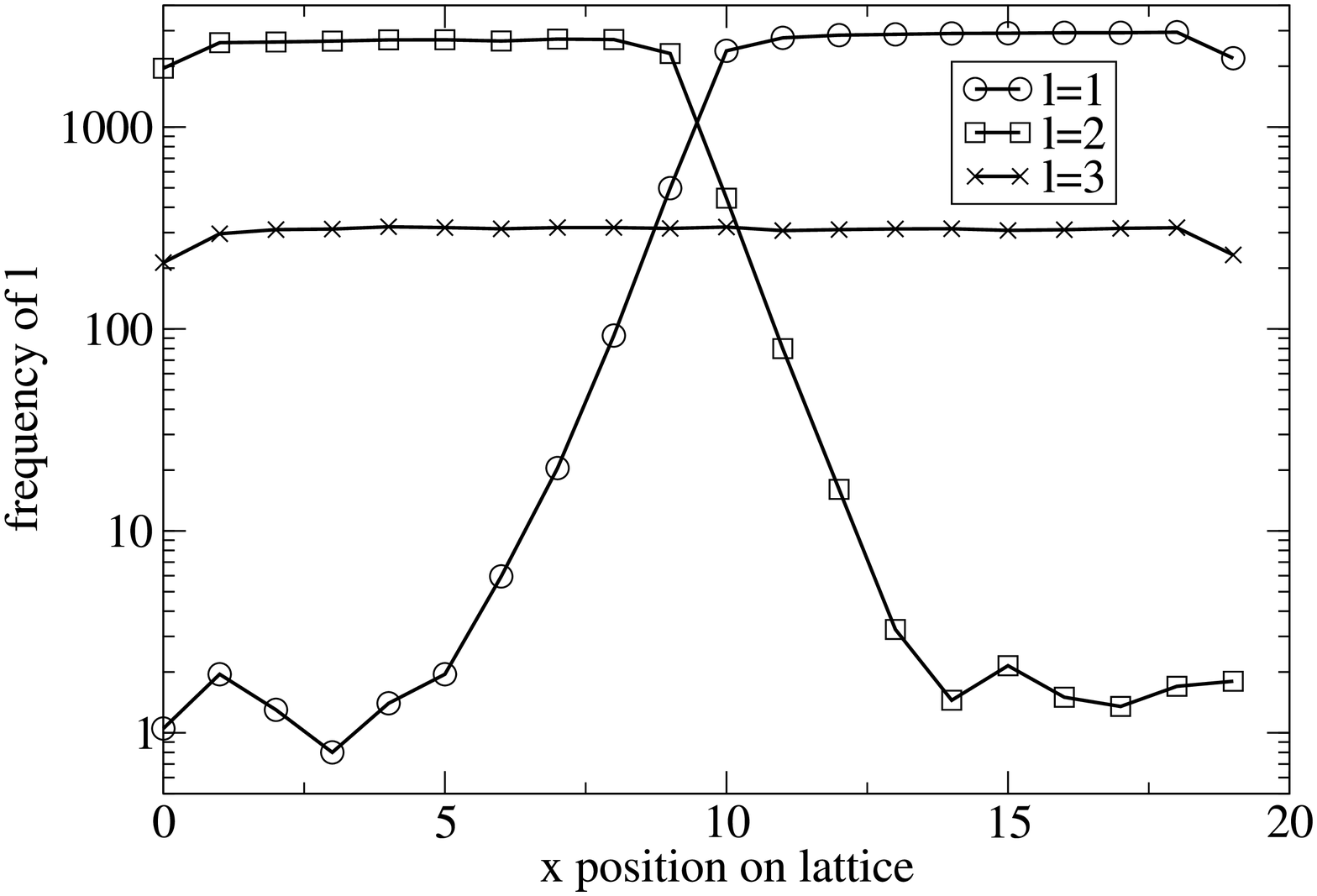}
\caption{Frequency for all $l$ on the interface. Bilinguals are homogeneously
	 distributed. The number of monolinguals decays exponentially at the interface.}
    \label{fig:distrB}
\end{center}
\end{figure}

We increased the number of agents by setting $N_{max} = 10,000,000$ and the initial 
populations to $100,000$ females and $100,000$ males for $p_f=1$: Now the 
exponential decay at the interface is observed clearly, as seen in 
Figure~\ref{fig:distrB}.

In our simulations we have also changed the parameter $d$, defining the number 
of neighbors an agent with $l=3$ examines, in order to know which language 
it can forget. The results for large $d$ are the same as for $d=1$.

\begin{figure}[htb]
  \begin{center}
    \includegraphics[width=0.8\textwidth]{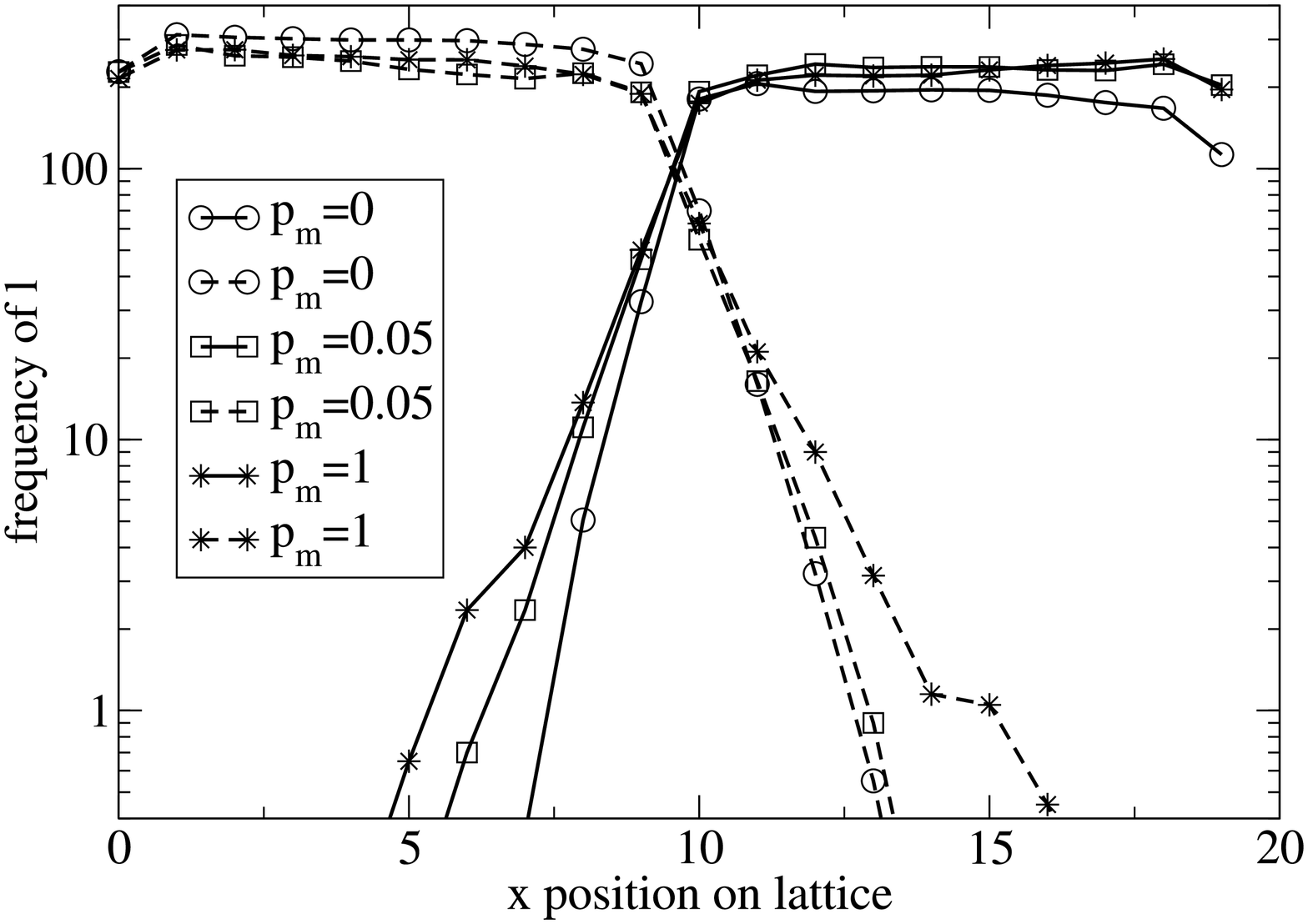}
\caption{Frequency of monolinguals on the lattice for different movement 
	 rates $p_m$ with $d=1$ and $p_f=0.2$.
 	 The larger $p_m$ is, the smoother becomes the steepness of the 
	 exponential decay.}
    \label{fig:distrM}
\end{center}
\end{figure}

The distribution of speakers on the lattice for different movement rates $p_m$ is
shown in Figure~\ref{fig:distrM}. We set $p_f=0.2$ and $d=1$.
Higher movement rates lead to a smoother interface.
Thus the exponential decay is weaker for large $p_m$.
At very high movement rates the interface becomes unstable.

\section{Conclusions}
\label{sec:concl}

We present simulations where the population of speakers of two 
different languages are of similar size for at least $10,000$ time steps. 
This meta-stable state is obtained only for a large number of agents per site 
and initialization of the lattice by distributing the speakers of different
languages on different halfs of the lattice. We can interpret that an interface of 
speakers in high populated areas, for instance at the Canal Street of New 
Orleans where on one side French and on the other side English is spoken,  
is more stable than in low populated areas. Different languages cannot survive 
for long times if their speakers are not geographically separated. 
Another result of the model is that the fraction of bilinguals relative to
monolinguals decreases exponentially with age. Older people, living for long time 
at the same place, do not need a second language. The mean value of the number of 
bilinguals versus the parameter $p_f$ to forget a second language shows a power law
with exponent $-1$.

The results of ref.~\cite{Schulze2005,Schulze2005b} are well reproduced in our 
model: At the interface the number of monolinguals decreases exponentially. 
Surprisingly, the number of bilinguals distributes rather homogeneously over 
the whole grid. The exponential decay of the number of monolinguals 
becomes steeper for smaller movement rates but is left unaltered by the 
lattice size. 
Higher movement rates lead to a more homogeneous distribution and can break the 
meta-stability of the interface. Nowadays, globalization gives us the possibility
to travel frequently over long distances and to stay larger periods at 
different places on Earth, one of the reasons why languages are becoming extinct.

We presented here the first model for language competition including 
ageing and sexual reproduction and reproduced well the results of other 
models although they are quite different. Numerical agent-based models 
on the computer yield interesting results despite
their simplicity, and we think that there will be much more to be done in future.

\section*{Acknowledgements}
I am funded by the DAAD (Deutscher Akademischer Austauschdienst) and 
thank D. Stauffer and S. Moss de Oliveira 
for very useful comments on my manuscript.

\end{document}